\documentclass{article}
\usepackage{spconf,amsmath,graphicx}

\usepackage{booktabs}
\usepackage{mathtools}
\usepackage{amssymb}

\DeclarePairedDelimiter{\norm}{\lVert}{\rVert}

\title{BWSNet: Automatic Perceptual Assessment Of Audio Signals}
\name{Clément Le Moine Veillon$^{\star}$$^{1}$, Victor Rosi$^{\star}$$^{2}$, Pablo Arias Sarah$^{3}$, Léane Salais$^{1}$, Nicolas Obin$^{1}$}
\address{$^{1}$STMS Lab - IRCAM, CNRS, Sorbonne Université, Paris, France \\
        $^{2}$Department of Speech Hearing and Phonetic Sciences, University College London, London, UK \\
        $^{3}$School of Psychology and Neuroscience, University of Glasgow, Glasgow, UK}

\begin{document}
%\ninept
%
\maketitle
\begin{abstract}

\end{abstract}

This paper introduces BWSNet, a model that can be trained from raw human judgements obtained through a Best-Worst scaling (BWS) experiment. It maps sound samples into an embedded space that represents the perception of a studied attribute. To this end, we propose a set of cost functions and constraints, interpreting trial-wise ordinal relations as distance comparisons in a metric learning task. We tested our proposal on data from two BWS studies investigating the perception of speech social attitudes and timbral qualities. For both datasets, our results show that the structure of the latent space is faithful to human judgements. 

\section{Introduction}
\label{sec:intro}

Access to a perceptual representation of data typically involves an experiment in which stimuli are judged by human participants according to a specific criterion. In particular, several methods to subjectively assess audio stimuli have been proposed, including pairwise comparison, MUSHRA and rating scales, the latter being widely praised in experimental psychology. With the advent of synthesis algorithms and the subsequent need to finely assess their outputs, such rating scales are widely employed to complement, or even replace, objective measures such as MCD or RMSE that are not necessarily correlated with human perception \cite{Theis2016}. Typically, the Mean Opinion Score (MOS) has been used to assess speech synthesis models' performance by asking participants to rate the quality or naturalness of output samples \cite{Zhang2019tts} and their similarity to a reference, according to specific speech attributes, e.g., speaker identity \cite{Zhang2019tts, Kameoka2020s2s}, emotion \cite{Zhou2022}, or attitude \cite{LeMoine2021VC}. Although favoured, scale-based methods present potential biases \cite{baumgartner2001, schuman1996}, making the choice of evaluation method an important issue in experimental psychology. 

Recently, attention was given to Best-Worst Scaling (BWS) \cite{Louviere2015}, another method in which participants are presented with trials of $N$ (e.g. $N$=4,5) items and asked to judge which ones are the best and worst according to a studied attribute. Once the experiment completed, each item receives a score - computed using more or less elaborate techniques on the basis of raw judgements - representing how it is perceived in regards with this attribute. BWS has proven to yield more reliable results than rating scales to gather perceptual scores \cite{Kiritchenko2017, hollis2018, RosiBWS} and has been found effective for various audio-related tasks such as the perception of speech emotions \cite{Zhou2022} and attitudes \cite{LeMoineThesis} as well as timbral qualities \cite{Rosi2023}. 

Unfortunately, all these subjective assessment methods always require a large number of human ratings to be reliable, making them costly and time-consuming. Facing this challenge, an entire field of research has emerged with the aim of automating perceptual evaluation. Thus, several methods have been proposed to learn a regression model that predicts MOS scores such as AutoMOS \cite{Patton2016}, Quality-Net \cite{Fu2018}, and MOSNet \cite{Lo2019, Choi2020, Choi2021}. To our knowledge and despite its aforementioned benefits, there is no existing method for predicting BWS judgements. 

In this paper, we introduce BWSNet, a model for automatic perceptual assessment based on BWS data. In contrast with existing MOSNet approaches, we do not seek to predict perceptual scores in a regression task. Indeed, the determination of BWS scores entails a dimensional reduction of the perceptual space underlying raw judgements, which involves a potential loss of information. To predict these judgements, that infer ordinal relations between items in each trial, we design a metric learning task in which these relations (see Fig. \ref{trial}) are interpreted as distances comparisons. BWSNet is thus trained to learn a function that maps sound samples into a latent space in which the distance represents samples dissimilarity with respect to the studied attribute. %We expect to use BWSNet to assess the perception of unseen samples, by examining the position of their embeddings in the latent space.

We present two contributions, firstly BWSNet, which consists of a set of cost functions adapted to the ordinal and relative nature of BWS judgements. Secondly, we apply it in the specific context of two studies previously led by the authors, investigating the perception of speech social attitudes \cite{LeMoineThesis} and timbral concepts \cite{Rosi2023}.

\vspace*{-\baselineskip}
\vspace{1mm}

\section{BWSNet}
\label{sec:BWS-Net}

\subsection{Problem Positioning}

A BWS trial is a tuple of $N$ sounds $t^{a} = \{x^{1}, ..., x^{N}\}$  set for judgement. For each trial, a judgment consists of choosing the best and worst items in the tuple, the $N-2$ remaining items are then considered neutrals. We denote $\mathcal{T}^{a}$ the set containing all the trials considered for the BWS experiment that investigates $a$.

\begin{figure}[h!]
    \centering
    \includegraphics[width=0.36\textwidth]{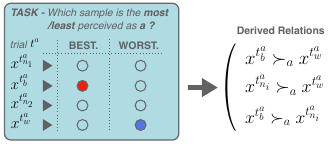}
    \caption{A BWS trial $t^{a} \in \mathcal{T}^{a}$ of $N=4$ sounds judged with respect to the attribute $a$ (left) and the derived relations (right).}
    \label{trial}
\end{figure}

Each sample is renamed with respect to the trial it lies in and the judgement it has been assigned. The best, worst and neutrals of trial $t^{a}$ can be indexed as $t_{b}^{a}$, $t_{w}^{a}$ and $t_{n_{i}}^{a}$ with $i \in  \{1, N-2\}$, respectively. We denote $\succ_{a}$, the relation such that $x \succ_{a} y$ is equivalent to  "$x$ is more perceived as $a$ than $y$". Then, $t^{a}$ judgement is represented by $2(N-1)$ relations - ordinal in nature - between sample triplets, as shown on the right part of Fig. \ref{trial}.

\vspace*{-\baselineskip}
\vspace{1mm}

\subsection{Concept of a BWSNet}

\begin{figure*}[h!]
    \centering
    \includegraphics[width=\textwidth]{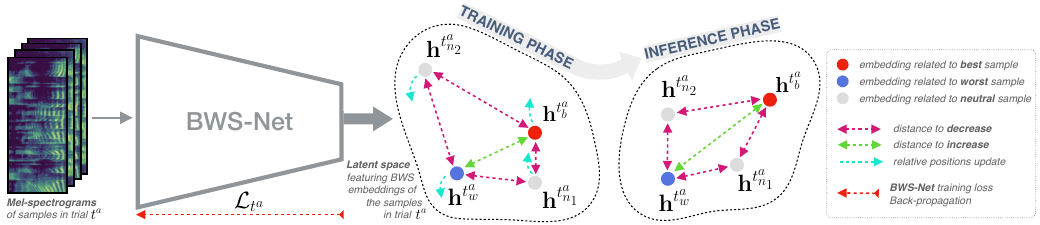}
    \caption{The Mel-Spectrograms of the $N=4$ samples related to a BWS trial $t^{a}$ investigating the attribute $a$ are passed to BWSNet. The model yields BWS embeddings of which relative position is changed over training.}
    \label{BWS-Net}
\end{figure*}

Our goal is to design a model able to learn perceptual representations underlying human BWS judgements. We use mel-spectrograms as sound representations, as it has proven to account for many perceptually relevant aspects of speech. As shown in Fig. \ref{BWS-Net}, for any sample $x$, the BWSNet takes a mel-spectrogram  $\mathbf{X}$ as input and produce a BWS embedding  $\mathbf{h}^{x}$. Relative positioning of embeddings, in the latent space they lie in, must be true to BWS judgments. To achieve this, ordinal relations between sound samples in BWS trials are translated into distances comparisons of the corresponding embeddings. Thus, introducing a distance $\lVert . \lVert : \mathbb{R}^{d} \rightarrow \mathbb{R}$, the relations for a trial $t^{a}$ displayed in Fig. \ref{trial} can be translated into the following inequalities for $i \in  \{1, ..., N-2 \}$:

\begin{align}
    \lVert \mathbf{h}^{t_{b}^{a}} - \mathbf{h}^{t_{w}^{a}} & \rVert \geq \lVert \mathbf{h}^{t_{b}^{a}} - \mathbf{h}^{t_{n_{i}}^{a}} \rVert \label{eq:ineq1} \\
    \lVert \mathbf{h}^{t_{b}^{a}} - \mathbf{h}^{t_{w}^{a}} & \rVert \geq \lVert \mathbf{h}^{t_{w}^{a}} - \mathbf{h}^{t_{n_{i}}^{a}} \rVert \label{eq:ineq2}
\end{align}

\subsection{Losses and Optimisation}

To train the model to match trials relations within its latent space, we designed our own training criterion, a cost function inspired by the metric learning literature and notably the triplet loss proposed in \cite{Hoffer2015}.

The relations to match are ordinal and, like in triplet loss, each one involves three items in a trial as formalized in inequalities \ref{eq:ineq1} and \ref{eq:ineq2}. However, they only prevail within a given trial, making the problem even more complex. To avoid mode collapse, i.e., the model turning all samples into one single point in the latent space, aforementioned inequalities must be considered strict. We thus introduce a positive margin $\alpha$ and define the Relative Contrastive (RC) loss $\mathcal{L}_{t^{a}}$ for any trial $t^{a} \in \mathcal{T}^{a}$ as:

\vspace*{-\baselineskip}
\vspace{0.8mm}

\begin{equation}
\small
    \begin{split}
    \mathcal{L}_{rc}^{t^{a}} =  \frac{1}{n_{v}^{t^{a}}} \sum_{i=1}^{N-2} & \max{(\norm{ \mathbf{h}^{t_{b}^{a}} - \mathbf{h}^{t_{n_{i}}^{a}} } - \norm{ \mathbf{h}^{t_{b}^{a}} - \mathbf{h}^{t_{w}^{a}} } + \alpha, 0)}
                + \\ 
                \frac{1}{n_{v}^{t^{a}}} \sum_{i=1}^{N-2} & \max{(\norm{ \mathbf{h}^{t_{w}^{a}} - \mathbf{h}^{t_{n_{i}}^{a}} } - \norm{ \mathbf{h}^{t_{b}^{a}} - \mathbf{h}^{t_{w}^{a}} } + \alpha, 0)} 
    \end{split}
\label{RC_loss}
\end{equation}

where $n_{v}^{t^{a}}$ is the number of relations that remain to be fulfilled within trial $t^{a}$.  

Perceptual differences represented in BWS trials can be more or less substantial, quantifying them requires dynamic margins rather than fixed ones. We thus introduce a network $\mathcal{M}$ dedicated to margin learning. It takes two parameters as arguments, a mean $\mu$ and an amplitude $\delta$ value, such that learnt margin lies between $\mu - \delta$ and $\mu + \delta$. With all trials' embeddings in the batch as input, it produces $N-2$ distinct margins $\{\alpha_{b,n_{i}}, \alpha_{w, n_{i}}\}_{i \in  \{ 1, N-2  \}}$ related to each trial relation. This impacts the RC loss formulated in equation \ref{RC_loss} as it takes learnt margins as additional input. Thus, for a given trial $t^{a}$, the Dynamic margin (Dm)-RC loss can be expressed as follows:  

\vspace*{-\baselineskip}
\vspace{0.8mm}

\begin{equation}
\small
\begin{split}
    \mathcal{L}_{drc}^{t^{a}} =  \frac{1}{n_{v}^{t^{a}}} \sum_{i=1}^{N-2} & \max{(\norm{ \mathbf{h}^{t_{b}^{a}} - \mathbf{h}^{t_{n_{i}}^{a}}} - \norm{ \mathbf{h}^{t_{b}^{a}} - \mathbf{h}^{t_{w}^{a}} } + \alpha_{b,n_{i}}, 0)}
                + \\
                \frac{1}{n_{v}^{t^{a}}} \sum_{i=1}^{N-2} & \max{(\norm{ \mathbf{h}^{t_{w}^{a}} - \mathbf{h}^{t_{n_{i}}^{a}} } - \norm{ \mathbf{h}^{t_{b}^{a}} - \mathbf{h}^{t_{w}^{a}} } + \alpha_{w,n_{i}}, 0)} 
\end{split}
\label{RCdm_loss}
\end{equation}

\vspace*{-\baselineskip}
\vspace{0.8mm}

We assume a Gaussian distribution of margins and propose an additional constraint to penalize our model's tendency to learn low-value margins. This constraint is formalized through a function $\gamma$ that takes the learned margins as an argument and outputs a scalar loss. Various functions can be tested, resulting in different learnt margin distributions. For a trial $t^{a}$, the Dynamic Margin Constraint (DMC) can be formulated as: 

\vspace*{-\baselineskip}
\vspace{0.8mm}

\begin{equation}
    \mathcal{L}_{dmc}^{t^{a}} = \sum_{i=1}^{N-2} \gamma(\alpha_{b, n_{i}} - \mu) + \gamma(\alpha_{w, n_{i}} - \mu)
\end{equation}

Decrease in the Dm-RC loss does not guarantee an increase in the number of fulfilled relations. Since margins can decrease overall without affecting order relationships. To avoid this, we have added a final loss directly derived from the Dm-RC loss which accounts, for a given trial $t^{a}$, to the number of unfulfilled relations within the trial divided by the number $N$ of elements in the trial.

\vspace*{-\baselineskip}
\vspace{0.8mm}

\begin{align}
    \mathcal{L}_{fr}^{t^{a}} = \frac{n_{v}^{t^{a}}}{N}
    \label{FR_loss}
\end{align}

\vspace*{-\baselineskip}
\vspace{0.8mm}

We introduce the scalars $\lambda_{dmc}$, $\lambda_{fr} \ge 0$ and define the global BWSNet loss for a trial $t^{a}$ as:

\vspace*{-\baselineskip}
\vspace{0.8mm}

\begin{align}
    \mathcal{L}^{t^{a}} = \mathcal{L}_{dmrc}^{t^{a}} + \lambda_{dmc} \mathcal{L}_{dmc}^{t^{a}} + \lambda_{fr}\mathcal{L}_{fr}^{t^{a}}
    \label{RC-DM_loss}
\end{align}

\section{Experiments}
\label{sec:typestyle}

We used data from two previous BWS studies to train and test our model. We provide short descriptions of these in the following. 

%\vspace*{-\baselineskip}
%\vspace{1.4mm}

\subsection{Data for Experiment}

\subsubsection{Study I: Speech Social Attitude}

%\noindent \textbf{\textsc{Experiment I: Speech Social Attitude.}} 
We carried out this first study on Att-HACK \cite{LeMoine2020}, a 30-hour speech dataset composed of 9 male and 11 female actors portraying four different social attitudes - \textit{friendliness}, \textit{dominance}, \textit{distance} and \textit{seduction} - over 100 sentences in French. 96 participants (48 women) were recruited to evaluate the perception of its related sounds for each a priori attitude, i.e. the ones already produced with aim of conveying this specific attitude. Model trainings were thus conducted on the four attitudes separately. For the sake of feasibility, only a fourth of the dataset has been assessed with each sound being evaluated eight times. 

\vspace*{-\baselineskip}
\vspace{0.8mm}

\subsubsection{Study II: Instrumental Timbre}

With the purpose of giving acoustic portraits of timbre concepts, sound expert participants (N=16, sound engineers and conductors) evaluated a dataset of musical instruments (N=520) on four well-known timbral concepts, namely \textit{brightness}, \textit{warmth}, \textit{roundness} and \textit{roughness}. The sounds showcase a great diversity of instruments, playing techniques, dynamics and registers. Unlike Study I, each sample is evaluated with respect to all concepts.

\vspace*{-\baselineskip}
\vspace{0.8mm}

\subsection{Implementation Details}

\subsubsection{Input Pipeline}

Mel-Spectrograms are obtained through computing Short-Term-Fourier-Transform (STFT) with FFT $2048$, hop $200$, window $800$ and $80$ mel filters. Our custom RC loss takes two tensors as input, in addition to the BWS embeddings: a tensor made of the corresponding trial names - which ensures the loss is computed trial-wise even if several trials lie in a single batch - and a tensor made of each element's corresponding judgement labels ($b$, $w$, $n$).

\vspace*{-\baselineskip}
\vspace{1mm}

\subsubsection{Architecture Design}

The BWSNet architecture was inspired by a promising version of ACRNN \cite{Li2019} initially proposed for speech emotion recognition. We previously assessed the role of its various components for speech attitude recognition in \cite{LeMoineThesis} using the same attitudinal dataset \cite{LeMoine2020}. Based on this study, the best configuration used $2$ convolutional layers with $64$ filters, temporal and feature kernel of sizes $5$ and $3$ respectively, and an $8$ head attention mechanism with $512$ dimensions. Here, we reuse the same architecture for our task - which is close to the one for which it has proved optimal - without further experimenting. Finer adjustments to the architecture, e.g. latent space dimension $d$, could improve performance depending on the input data, especially for timbral concepts. Here we choose $d=32$ and focus on the relative importance of the aforementioned cost functions and constraints.

\vspace*{-\baselineskip}
\vspace{1mm}

\subsubsection{Training Procedure}

In both experiments, we split data into three groups, first we selected $10\%$ of the samples and dropped out any trial they were involved in. Then, we split the remaining data trial-wise into training ($80\%$) and validation ($20\%$) trial sets.  We fed our model with batches of size $80$ with as many data of each concept for Experiment II and used ADAM optimizer with $0.0001$ as learning rate. We found the euclidean distance for $\lVert . \lVert$, $\gamma : x \rightarrow ReLU(-x)$ for DMC and $\mu = \gamma = 1$ as margin learning module $\mathcal{M}$ parameters to yield the best results.%All codes are written in Python-Tensorflow 2.8.

\vspace*{-\baselineskip}
\vspace{1mm}

\subsubsection{Evaluation Criteria}

To evaluate the performance of BWSNet, we used two metrics, reflecting the arrangement of speech samples in the latent space at two levels: \textbf{FR} and \textbf{WAT} - respectively the percentages of fulfilled relations and well-arranged trials within the set - both computed on relations from dropped trial set, involving at least one unseen sample. We chose lowest \textbf{FR} value as early stopping criterion.

\section{Results \& Discussion}

\vspace*{-\baselineskip}
\vspace{4mm}

We carried out an ablation study to demonstrate the influence of each cost function on BWSNet's performance. Then, we sought to explore the latent spaces it yielded for both studies.

\vspace*{-\baselineskip}
\vspace{2mm}

\subsection{Ablation Study}

Table \ref{results} displays \textbf{FR} and \textbf{WAT} mean and standard deviation values for various BWSNet configurations evaluated on unseen samples across both BWS studies' sets of attributes. First, as expected, the fixed-margin configuration (\textbf{A-f}) did not generalize well, as some trials' best and worst could be either very distant in the latent space or rather close. Then, when considering learnt margin configurations, it appeared that with no constraint on margins (\textbf{A-l}) the latent space was collapsing into one single point, turning distances between any pair of points null. Adding such a constraint (\textbf{A-l-d}) tended to help the model converge and fulfill just under one relations in two on attitudes and slightly more than one relations in two for timbre, which is insufficient to claim that our model accurately predicts BWS judgements. However, we found that BWSNet alternatively uses two strategies to lower Dm-RC loss: it could seek to fulfill more relations within trials otherwise it could reduce the margins. The addition of FR loss as training criterion (\textbf{A-l-d-fr}) appeared to prevent the model from engaging in the second strategy and led to improvements by $27.7\%$ and $27.5\%$ in both \textbf{FR} and \textbf{WAT} respectively for attitudes and by $4.7\%$ in \textbf{FR} for timbre. The model performed better for attitudinal than for timbral data, which may be due to the choice of neural architecture that we deliberately adapted to the former.

\begin{table}[h!]
\centering
\resizebox{\columnwidth}{!}{
\begin{tabular}{lcccccc}
\toprule

  &   &   & \multicolumn{2}{c}{\textit{Study I: Attitudes}} & \multicolumn{2}{c}{\textit{Study II: Instrument Timbre}} \\  
\textbf{Model} & $\lambda_{dmc}$ & $\lambda_{fr}$ & \textbf{FR} (\%)   & \textbf{WAT} (\%)  & \textbf{FR} (\%) & \textbf{WAT} (\%) \\
\cmidrule(lr){1-3} \cmidrule(lr){4-5} \cmidrule(lr){6-7}
\textbf{A-f}      & -  & - & $21.4 \pm 8.5$  & $5.8 \pm 3.6$   & $37.5 \pm 2.5$  &  $4.7 \pm 0.5$  \\
\textbf{A-l}      & 0 & 0 &  $1.0 \pm 0.4$   &  $0.2 \pm 0.0$  & $26.1 \pm 2.1$  &     $\textbf{26.0} \pm \textbf{1.3}$ \\
\textbf{A-l-d}    & 1 & 0 &  $40.1 \pm 18.7$ & $22.4 \pm 17.1$ & $51.6 \pm 3.5$  & $19.9 \pm 2.6$ \\
\textbf{A-l-d-fr} & 1 & 1 &  $\textbf{67.7} \pm \textbf{4.5}$  & $\textbf{49.9} \pm \textbf{8.9}$  & $\textbf{56.3} \pm \textbf{2.4}$  &  $23.9 \pm 1.2$  \\
\bottomrule
\end{tabular}}

\caption{\textbf{FR} \& \textbf{WAT} mean and standard deviation values for various BWSNet configurations evaluated on unseen samples across both BWS studies' sets of attributes.}
\label{results}
\end{table}

\vspace*{-\baselineskip}
\vspace{-1.2mm}

\subsection{Exploring BWSNet's Latent Space}

\begin{figure}[t]
    \centering
    \includegraphics[width=0.49\textwidth]{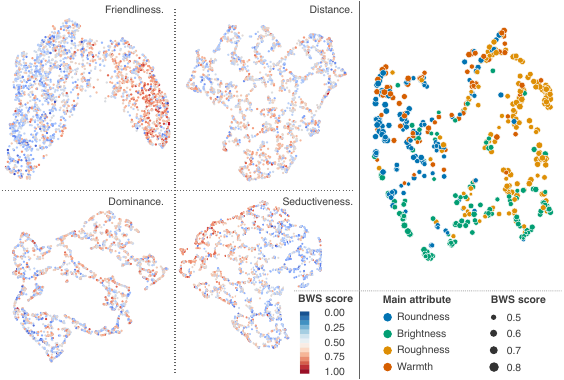}
 \caption{BWSNet's latent space UMAP vizualization for social attitudes (left) and timbral qualities (right). Each point is a sample whose BWS score is represented by its colour (left) and size (right) respectively.}
    \label{latent}
\vspace*{-\baselineskip}
\end{figure}

To further analyse our results, we investigated how BWS scores, obtained with the original scoring algorithm \cite{hollis2018}, relate to the dimensions of our latent space. These scores ranging from 0 (poor) to 1 (excellent) reflect the perceived correspondence of items' with a given attribute. Figure \ref{latent} shows the BWSNet latent spaces corresponding to each attitude (left) and timbre qualities (right) with the original scores characterised with color and size respectively.

By assessing attitudes independently in the BWS experiment, we obtained four distinct latent spaces with different degrees of polarisation with regard to BWS scores. For instance, for \textit{friendliness} and \textit{seductiveness} - that were found the most consensual \cite{LeMoineThesis} - high and low score samples are located in opposite ends of the space. Furthermore, regarding the latent spaces obtained for \textit{distance} and \textit{dominance}, BWS scores may be an inaccurate depiction of their multi-dimensional nature. 

As for timbral concepts, the original study sought to uncover their mutual interactions which leads to have all samples lying in the same space. To report on these interactions, we associated each sound with its most salient attribute (i.e., the highest scored concept for each sound). We observe similar interactions to those in the original study, e.g. \textit{warmth} and \textit{roundness} are blended together in the timbral space while showing strong opposition to \textit{brighthness}. 

%\vspace*{-\baselineskip}
%\vspace{1mm}

\vspace*{-\baselineskip}
\vspace{2mm}

%\vspace*{-\baselineskip}
%\vspace{1.2mm}

\section{Conclusion}
\label{sec:majhead}

\vspace*{-\baselineskip}
\vspace{4mm}

This paper presents BWSNet, a model dedicated to automatic audio perceptual assessment based on BWS judgements. Distances between learnt sample embeddings in the resulting latent space represent their perceptual similarity. By fulfilling almost 70\% of relations involving an unseen sample for attitudinal speech data, BWSNet provides a rather accurate estimation of how this sample is perceived based on its distance with previously judged samples in the latent space. Its performance on timbral data (56\% fulfilled relations) also indicates potential for application to a manifold of judgement tasks. These results, obtained on two very different datasets, mark a first step towards automating the BWS perceptual assessment. Furthermore, as the analysis of its latent space suggests, BWSNet could be a relevant tool for representing and understanding human perception. Provided that participants agree to a certain extent, using more BWS judgements for training would likely yield a more comprehensive map of the perceptual space for a chosen sound attribute.

\vfill\pagebreak

\bibliographystyle{IEEEbib}
\bibliography{ref}%,

\begin{thebibliography}{10}

\bibitem{Theis2016}
Lucas Theis, A{\"a}ron van~den Oord, and Matthias Bethge,
\newblock ``A note on the evaluation of generative models,''
\newblock {\em arXiv preprint arXiv:1511.01844}, 2016.

\bibitem{Zhang2019tts}
Mingyang Zhang, Xin Wang, Fuming Fang, Haizhou Li, and Junichi Yamagishi,
\newblock ``Joint training framework for text-to-speech and voice conversion using multi-source tacotron and wavenet,''
\newblock {\em arXiv preprint arXiv:1903.12389}, 2019.

\bibitem{Kameoka2020s2s}
Hirokazu Kameoka, Kou Tanaka, Damian Kwa{\'s}ny, Takuhiro Kaneko, and Nobukatsu Hojo,
\newblock ``Convs2s-vc: Fully convolutional sequence-to-sequence voice conversion,''
\newblock {\em IEEE/ACM Transactions on audio, speech, and language processing}, vol. 28, pp. 1849--1863, 2020.

\bibitem{Zhou2022}
Kun Zhou, Berrak Sisman, Rajib Rana, Bj{\"o}rn~W Schuller, and Haizhou Li,
\newblock ``Speech synthesis with mixed emotions,''
\newblock {\em IEEE Transactions on Affective Computing}, 2022.

\bibitem{LeMoine2021VC}
Cl{\'e}ment Le~Moine, Nicolas Obin, and Axel Roebel,
\newblock ``{Towards end-to-end F0 voice conversion based on Dual-GAN with convolutional wavelet kernels},''
\newblock in {\em {EUSIPCO}}, Dublin (virtual ), Ireland, 2021.

\bibitem{baumgartner2001}
Hans Baumgartner and Jan-Benedict~EM Steenkamp,
\newblock ``Response styles in marketing research: A cross-national investigation,''
\newblock {\em Journal of marketing research}, vol. 38, no. 2, pp. 143--156, 2001.

\bibitem{schuman1996}
Howard Schuman and Stanley Presser,
\newblock {\em Questions and answers in attitude surveys: Experiments on question form, wording, and context},
\newblock Sage, 1996.

\bibitem{Louviere2015}
Jordan~J. Louviere, Terry~N. Flynn, and A.~A.~J. Marley,
\newblock {\em Best-Worst Scaling: Theory, Methods and Applications},
\newblock Cambridge University Press, 2015.

\bibitem{Kiritchenko2017}
Svetlana Kiritchenko and Saif Mohammad,
\newblock ``Best-worst scaling more reliable than rating scales: A case study on sentiment intensity annotation,''
\newblock in {\em Proceedings of the 55th Annual Meeting of the Association for Computational Linguistics (Volume 2: Short Papers)}, Vancouver, Canada, July 2017, pp. 465--470, Association for Computational Linguistics.

\bibitem{hollis2018}
Geoff Hollis,
\newblock ``Scoring best-worst data in unbalanced many-item designs, with applications to crowdsourcing semantic judgments,''
\newblock {\em Behavior research methods}, vol. 50, no. 2, pp. 711--729, 2018.

\bibitem{RosiBWS}
Victor Rosi, Aliette Ravillion, Olivier Houix, and Patrick Susini,
\newblock ``Best-worst scaling, an alternative method to assess perceptual sound qualities,''
\newblock {\em The Journal of the Acoustical Society of America}, vol. 2, pp. 064404, 06 2022.

\bibitem{LeMoineThesis}
Clément Le~Moine~Veillon,
\newblock {\em Neural Conversion of Social Attitudes in Speech Signals},
\newblock Ph.D. thesis, 2023,
\newblock Thèse de doctorat dirigée par Roebel, Axel et Obin, Nicolas Informatique Sorbonne université 2023.

\bibitem{Rosi2023}
Victor Rosi, Pablo Arias~Sarah, Olivier Houix, Nicolas Misdariis, and Patrick Susini,
\newblock ``Shared mental representations underlie metaphorical sound concepts,''
\newblock {\em Scientific Reports}, vol. 13, no. 1, pp. 5180, 2023.

\bibitem{Patton2016}
Brian Patton, Yannis Agiomyrgiannakis, Michael Terry, Kevin Wilson, Rif~A. Saurous, and D.~Sculley,
\newblock ``Automos: Learning a non-intrusive assessor of naturalness-of-speech,''
\newblock in {\em NIPS 2016 End-to-end Learning for Speech and Audio Processing Workshop}, 2016.

\bibitem{Fu2018}
Szu-Wei Fu, Yu~Tsao, Hsin-Te Hwang, and Hsin-min Wang,
\newblock ``Quality-net: An end-to-end non-intrusive speech quality assessment model based on {BLSTM},''
\newblock 09 2018, pp. 1873--1877.

\bibitem{Lo2019}
Chen-Chou Lo, Szu-Wei Fu, Wen-Chin Huang, Xin Wang, Junichi Yamagishi, Yu~Tsao, and Hsin-Min Wang,
\newblock ``{{MOSNet}: Deep Learning-Based Objective Assessment for Voice Conversion},''
\newblock in {\em Proc. Interspeech}, 2019, pp. 1541--1545.

\bibitem{Choi2020}
Yeunju Choi, Youngmoon Jung, and Hoirin Kim,
\newblock ``{Deep MOS Predictor for Synthetic Speech Using Cluster-Based Modeling},''
\newblock in {\em Proc. Interspeech}, 2020, pp. 1743--1747.

\bibitem{Choi2021}
Yeunju Choi, Youngmoon Jung, and Hoirin Kim,
\newblock ``Neural {MOS} prediction for synthesized speech using multi-task learning with spoofing detection and spoofing type classification,''
\newblock in {\em 2021 IEEE Spoken Language Technology Workshop (SLT)}, 2021, pp. 462--469.

\bibitem{Hoffer2015}
Elad Hoffer and Nir Ailon,
\newblock ``Deep metric learning using triplet network,''
\newblock in {\em Similarity-Based Pattern Recognition}, Aasa Feragen, Marcello Pelillo, and Marco Loog, Eds., Cham, 2015, pp. 84--92, Springer International Publishing.

\bibitem{LeMoine2020}
Cl{\'e}ment Le~Moine and Nicolas Obin,
\newblock ``{Att-HACK: An Expressive Speech Database with Social Attitudes},''
\newblock in {\em {Speech Prosody}}, Tokyo, Japan, 2020.

\bibitem{Li2019}
Yuanchao Li, Tianyu Zhao, and Tatsuya Kawahara,
\newblock ``{Improved End-to-End Speech Emotion Recognition Using Self Attention Mechanism and Multitask Learning},''
\newblock in {\em Proc. Interspeech}, 2019, pp. 2803--2807.

\end{thebibliography}
%\bibliography{strings,refs}

\end{document}